\documentclass[%
 reprint,
 amsmath,amssymb,
 aps,
]{revtex4-2}

\usepackage{graphicx}
\usepackage{dcolumn}
\usepackage{bm}
\usepackage[hidelinks]{hyperref}
\usepackage{xcolor}
\usepackage{mathtools}
\usepackage{tikz}

\definecolor{lime}{HTML}{A6CE39}
\newcommand{\orcidicon}{%
	\begin{tikzpicture}
	\draw[lime, fill=lime] (0,0) 
		circle [radius=0.12] 
		node[white] {{\fontfamily{qag}\selectfont \tiny ID}};
	\draw[white, fill=white] (-0.0625,0.095) 
		circle [radius=0.005];
	\end{tikzpicture}
	\hspace{-5mm}
}
\newcommand\orcidJosh{{\href{https://orcid.org/0000-0003-1200-7261}{\orcidicon}}}

\begin{document}
\newcommand{\red}[1]{{\color{red} #1}}
\newcommand{\blue}[1]{{\color{blue} #1}}


\title{A theory agnostic uniqueness theorem for the Kerr solution}

\author{Joshua Baines\!\orcidJosh\,}
 \email{joshua.baines@unimelb.edu.au}
\affiliation{University of Melbourne, School of Physics, Parkville, VIC 3010, Australia.}
\affiliation{Australian Research Council Centre of Excellence for Gravitational Wave Discovery \\
\null\qquad (OzGrav), University of Melbourne, Parkville, VIC 3010, Australia.}

\date{\today; \LaTeX-ed \today }

\begin{abstract}
In this paper, we show that under a suitable set of symmetry arguments and asymptotic conditions, the uniqueness of the Kerr solution can be proven. None of the conditions considered herein explicitly assume the validity of the Einstein equations. Hence, we are able to construct a theory agnostic uniqueness theorem for the Kerr spacetime. This result has implications for theories of quantum gravity or modified theories of gravity which wish to excise singular behaviour from their corresponding spacetimes. Furthermore, this work is complimentary to Penrose's singularity theorem. While the spacetime considered herein is not as general as that assumed by Penrose, we show that singularities can be unavoidable even when the validity of the Einstein equations is not presupposed.
\end{abstract}

\maketitle


\noindent\emph{Introduction:}

Since its discovery in 1963 \cite{Kerr:1963,Kerr2,Kerr3,Kerr4}, the Kerr solution has stood as the most widely accepted model for astrophysical black holes. It has been employed across an enormous range of astrophysical settings and subjected to extensive observational scrutiny, passing tests of general relativity in the weak field regime and, increasingly, in the strong field regime probed by gravitational wave detections \cite{Abbott:2016a,Abbott:2016b,LIGOScientific:2025} and very long baseline interferometry \cite{EHT:2019,EHT:2022} (see \cite{Will:2014,Berti:2015,Bambi:2017} for reviews). Despite this success, a substantial body of work has sought alternatives to the Kerr solution. These alternatives broadly fall into two classes: regular black holes, in which the curvature singularity is replaced by a non-singular core \cite{Bardeen:1968,AyonBeato:1998,Hayward:2006,Simpson:2019,Simpson:2021}, and horizonless black hole mimickers \cite{MazurMottola:2004,CardosoPani:2019}. Both are motivated, at least in part, by the desire to rid gravitational physics of curvature singularities.

These models can avoid singularities precisely because they do not assume the Einstein equations to hold. Consequently, one of the hypotheses of Penrose's singularity theorem \cite{Penrose:1965,HawkingPenrose:1970} is no longer satisfied, and the spacetime is not guaranteed to contain a singularity. The same loophole affects uniqueness, the classical uniqueness theorems for the Kerr solution \cite{Israel:1967,Carter:1971,Hawking:1972,Robinson:1975,Mazur:1982} (see \cite{Heusler:1996,Chrusciel:2012} for reviews) all require the vacuum field equations as an input. Once this requirement is dropped, one is seemingly free to abandon the Kerr solution in favour of alternative spacetimes. In this paper we show that this freedom is not always given. Even without enforcing the Einstein equations, the Kerr solution is recovered from a set of symmetry and asymptotic conditions alone. In this sense, the uniqueness result presented here is theory agnostic.

It can be argued that real astrophysical black holes are expected to satisfy a number of physically motivated requirements. An isolated black hole in equilibrium should be invariant under time translations and axial rotations. Hence the spacetime is stationary and axisymmetric. Furthermore, we have certainly observed long lived accretion disks around astrophysical black holes, in order to ensure the stability of such disks, we require the absence of chaotic particle orbits \cite{Apostolatos:2009}. This in turn requires that the Hamilton--Jacobi equation is timelike separable \cite{Carter:1968a,Carter:1968b,WalkerPenrose:1970}. This ensures the existence of a fourth constant of the motion beyond mass, energy, and axial angular momentum. Which in turn, makes the geodesic equation integrable. However, geodesic motion is only the leading description of test body dynamics. Realistic astrophysical sources, extreme and intermediate mass ratio inspirals \cite{Tanaka:1996,AmaroSeoane:2007,Poisson:2011,Steinhoff:2012}, involve compact spinning bodies whose motion is governed, to dipole order, by the Mathisson--Papapetrou--Dixon (MPD) equations \cite{Mathisson:1937,Papapetrou:1951,Dixon:1970}. This system is in general non-integrable and admits chaotic solutions even in Schwarzschild \cite{SuzukiMaeda:1997}. So the natural extension of Hamilton--Jacobi separability is then the requirement that linear in spin MPD motion be integrable, which is ensured by the existence of a Killing--Yano tensor \cite{Rudiger:1981,Witzany:2019}. The associated separability of the Klein--Gordon and Teukolsky equations \cite{Teukolsky:1973} governs the propagation of scalar and gravitational perturbations, and hence the quasinormal mode spectrum \cite{Kokkotas:1999,Berti:2009}, which is consistent with current LVK ringdown observations\cite{LIGOScientific:2025}. A spacetime modelling a real astrophysical black hole is also expected to be asymptotically flat, although not everywhere conformally flat, and, at large distances, should reproduce the gravitational potential of a point mass. Herein we show that this set of conditions is sufficient to establish the uniqueness of the Kerr solution. \\

\noindent\emph{Key result:}

We begin with a statement of the theorem.\\

\noindent\emph{Theorem:} 

Suppose a 4-dimensional spacetime satisfies the following conditions:
\begin{enumerate}
    \item The spacetime is stationary and axisymmetric
    \item The spacetime admits a timelike separable Hamilton--Jacobi equation 
    \item The spacetime admits a separable Klein--Gordon equation
    \item The spacetime admits a Killing--Yano tensor
    \item The spacetime admits a separable Teukolsky equation
    \item The spacetime is asymptotically flat
    \item The spacetime is not conformally flat
    \item The spacetime reproduces the Newtonian gravitational potential of a point mass at sufficiently large $r$
\end{enumerate}
Then the spacetime is uniquely given by the Kerr solution. \\

\noindent\emph{Proof:}

As shown in \cite{Papadopoulos:2020,Papadopoulos:2018} (based on work in \cite{Benenti:1979}) the inverse metric of a stationary axisymmetric spacetime in which the Hamilton--Jacobi equation is timelike separable can be written, in $(r,\theta,\phi,t)$ coordinates, as follows 
\begin{equation} \label{PK-10}
g^{ab}(r,\theta) = 
\frac{1}{A_1 + B_1}
\left[\begin{array}{cccc}
A_2 & 0 & 0 & 0\\ 
0 & B_2 & 0 & 0\\
0 & 0 & A_3+B_3 & \;A_4+B_4\\
0 & 0 & A_4+B_4 & \;A_5+B_5
\end{array}\right] ,
\end{equation}
where the $A_i$ functions are functions purely of $r$ and the $B_i$ functions are functions purely of $\theta$. Since the Hamilton--Jacobi equation is separable, the corresponding Killing tensor is
\begin{widetext}
\begin{equation} \label{E:Killing_tensor}
K^{ab}(r,\theta) = 
{1\over  A_1 + B_1}
\left[\begin{array}{cccc}
A_2B_1 & 0 & 0 & 0\\
0 & -B_2A_1 & 0 & 0\\
0 & 0 & B_1A_3-A_1B_3 &\; B_1A_4-A_1B_4\\
0 & 0 & B_1A_4-A_1B_4 & \;B_1A_5-A_1B_5
\end{array}\right] .
\end{equation}
\end{widetext}

Since the spacetime is asymptotically flat, it must reduce to the metric of flat space as $r\rightarrow\infty$. The inverse metric of flat space in oblate spheroidal coordinates is given, in $(t,r,\theta,\phi)$ coordinates, as
\begin{equation} \label{E:inv_metric_flat_OS}
g^{ab} = \left[\begin{array}{cccc}
    -1 & 0 & 0 & 0 \\
     0 & \frac{r^2+a^2}{\Sigma} & 0 & 0 \\
     0 & 0 & \frac{1}{\Sigma} & 0 \\
     0 & 0 & 0 & \frac{1}{(r^2+a^2)\sin^2\theta}
\end{array}\right] ,
\end{equation}
where $\Sigma = r^2+a^2\cos^2\theta$. If we compare this to equation \eqref{PK-10}, we see that 
\begin{equation} \label{E:Asym_flat_condition_1}
\frac{A_2}{A_1+B_1}=\frac{r^2+a^2}{r^2+a^2\cos^2\theta}\,;
\end{equation}
\begin{equation} \label{E:Asym_flat_condition_2}
\frac{B_2}{A_1+B_1} = \frac{1}{r^2+a^2\cos^2\theta}\,;
\end{equation}
\begin{equation} \label{E:Asym_flat_condition_3}
\frac{A_3+B_3}{A_1+B_1} = \frac{1}{(r^2+a^2)\sin^2\theta}\,;
\end{equation}
\begin{equation} \label{E:Asym_flat_condition_4}
\frac{A_5+B_5}{A_1+B_1} = -1;
\end{equation}
\begin{equation} \label{E:Asym_flat_condition_5}
\frac{A_4+B_4}{A_1+B_1} = 0.
\end{equation}
From equation \eqref{E:Asym_flat_condition_1} we can make the correspondence
\begin{equation}
B_1=a^2\cos^2\theta\,;\quad \lim_{r\rightarrow\infty} A_1 = r^2; \quad \lim_{r\rightarrow\infty} A_2 = r^2+a^2.
\end{equation}
Knowing this, we can infer from equation \eqref{E:Asym_flat_condition_2} that 
\begin{equation}
B_2=1.
\end{equation}
Then via equation \eqref{E:Asym_flat_condition_3} we have 
\begin{equation}
A_3+B_3 = \frac{A_1+B_1}{(r^2+a^2)\sin^2\theta} = \csc^2\theta - \frac{a^2}{r^2+a^2}\,,
\end{equation}
hence 
\begin{equation}
B_3 = \csc^2\theta\,; \quad \lim_{r\rightarrow\infty} A_3 = -\frac{a^2}{r^2+a^2}\,.
\end{equation}
From equation \eqref{E:Asym_flat_condition_4} we have 
\begin{widetext}
\begin{equation}
A_5+B_5 = -(A_1+B_1) = -(r^2+a^2\cos^2\theta)=-(r^2-a^2)+a^2\sin^2\theta,
\end{equation}
\end{widetext}
hence
\begin{equation}
B_5 = a^2\sin^2\theta\,;\quad \lim_{r\rightarrow\infty} A_5 = -(r^2-a^2).
\end{equation}
Lastly, from equation \eqref{E:Asym_flat_condition_5} we have 
\begin{equation}
\frac{A_4+B_4}{A_1+B1} = 0,
\end{equation}
that is
\begin{equation}
A_4+B_4 = 0,
\end{equation}
hence we have 
\begin{equation}
B_4 =c, \quad \lim_{r\rightarrow\infty} A_4 = -c,
\end{equation}
where $c$ is some constant. \\

Now we wish to enforce separability of the Klein--Gordon equation. It can be shown \cite{PGLT2,Franzin:2021,Commutator:2002,Giorgi:2021} (see also \cite{Carter:1968b,Teukolsky:1972}) that this condition is equivalent to enforcing
\begin{equation} \label{E:KG_condition}
\nabla_a[K,R]^a{}_b = \nabla_a(K^a{}_c\, R^c{}_b - R^a{}_c\,K^c{}_b) = 0.
\end{equation}
Using the metric and Killing tensor as shown in equations \eqref{PK-10} and \eqref{E:Killing_tensor} respectively, along with the $B_i$ functions as found above, we find 3 constraints
\begin{equation}
A_3 = \frac{a^2A_5}{(A_1+a^2)^2}\,;\quad
A_4 = \frac{a A_5}{A_1+a^2}\,;\quad
B_4 = a.
\end{equation}
With these conditions, equation \eqref{E:KG_condition} is satisfied. \\

Now we are left with 3 free functions $A_1$, $A_2$ and $A_5$. Let us define
\begin{equation}
A_1=\Xi(r)^2; \quad A_2=\Delta(r);
\end{equation}
and
\begin{equation}
A_5=-\frac{\exp(2\Phi(r))(\Xi(r)^2+a^2)^2}{\Delta(r)},
\end{equation}
then we are left with the metric
\begin{widetext}
\begin{equation} \label{3f_kerr_metric}
\begin{split}
ds^2= & -\frac{\Delta(r)\exp(-2\Phi(r))-a^2\sin^2\theta}{\Xi(r)^2+a^2\cos^2\theta}\; dt^2
+\frac{\Xi(r)^2+a^2\cos^2\theta}{\Delta(r)}\; dr^2+(\Xi(r)^2+a^2\cos^2\theta)\;d\theta^2\\
&-2\;\frac{a \sin^2\theta \;(\Xi(r)^2-\Delta(r)\exp(-2\Phi(r))+a^2)}{\Xi(r)^2+a^2\cos^2\theta}\;dtd\phi+\frac{ \left((\Xi(r)^2+a^2)^2-\exp(-2\Phi(r))\Delta(r)a^2\sin^2\theta\right)\sin^2\theta}{\Xi(r)^2+a^2\cos^2\theta}\; d\phi^2 .
\end{split}
\end{equation}
\end{widetext}
As shown in \cite{3-function}, the metric in equation \eqref{3f_kerr_metric} admits a Killing–Yano tensor (an antisymmetric rank 2 tensor which satisfies the Killing--Yano equation $\nabla_{(c}Y_{a)b}=0$) when $\Phi(r) = 0$ and $\Xi(r) = \pm r$. Since the spacetime depends on $\Xi(r)^2$ and not just $\Xi(r)$, we may choose $\Xi(r) = r$ without loss of generality.\\

The Teukolsky equation is a wave equation describing the propagation of linearised gravitational radiation. In particular, the equation can be used to describe gravitational waves emitted during the ringdown phase of binary black hole merger events which one can compare to LVK data as tests of general relativity. As shown in \cite{Li:2022pcy} we can write the Teukolsky equation in modified gravity via 
\begin{equation}
\left(\mathcal{E}^B_2F^B_2-\mathcal{E}^B_1F^B_1-2\Psi^B_2\right)\Psi^P_0=\mathcal{S}^P,
\end{equation}
where 
\begin{equation} \label{E:NP_operators}
\begin{gathered}
F_1 = \bar{\delta}-4\alpha+\pi;\\
F_2 = \boldsymbol{\Delta}-4\gamma+\mu;\\
\mathcal{E}_1 = \delta-\tau+\bar{\pi}-\bar{\alpha}-3\beta-\frac{1}{\Psi_2}\delta\Psi_2;\\
\mathcal{E}_2 = D-\rho-\bar{\rho}-3\epsilon+\bar{\epsilon}-\frac{1}{\Psi_2}D\Psi_2;\\
\end{gathered}
\end{equation}
and
\begin{equation}
\begin{gathered}
\mathcal{S}^P=\mathcal{E}_2^BS_2^P-\mathcal{E}_1^BS_1^P;\\
\begin{multlined}
S_1 = (\delta+\bar{\pi}-2\bar{\alpha}-2\beta)\Phi_{00}-(D-2\epsilon-2\bar{\rho})\Phi_{01}\\
+2\sigma\Phi_{10}-2\kappa\Phi_{11}-\bar{\kappa}\Phi_{02};
\end{multlined}
\\
\begin{multlined}
S_2 = (\delta+2\bar{\pi}-2\beta)\Phi_{01}-(D-2\epsilon+2\bar{\epsilon}-\bar{\rho})\Phi_{02}\\
-\bar{\lambda}\Phi_{00}+2\sigma\Phi_{11}-2\kappa\Phi_{12},
\end{multlined}
\end{gathered}
\end{equation}
which are functions of Newman--Penrose spin coefficients, directional derivatives, Ricci coefficients and the background Weyl scalar $\Psi^B_2$ given by
\begin{equation}
\Psi_2^B = \frac{(r-ia\cos\theta)^3}{12\Sigma}\frac{d^2}{dr^2}\left[\frac{\Delta(r)-r^2-a^2}{(r-ia\cos\theta)^3}\right],
\end{equation}
(see \cite{Li:2022pcy,Chandrasekhar:1985kt} for details). Where we have decomposed the metric into $g_{ab}=(g^B)_{ab}+\varepsilon (h^P)_{ab}$, where $\varepsilon$ is a small constant. That is, the $B$ stands for the background spacetime (i.e. equation \eqref{3f_kerr_metric} with $\Phi(r)=0$ and $\Xi(r)=r$) and the $P$ stands for the perturbation (i.e. the outgoing gravitational wave). In order to give the Teukolsky equation in terms of coordinates, we require specification of the directional derivatives and the Newman--Penrose spin coefficients shown in equation \eqref{E:NP_operators}, this then requires specification of a suitable null tetrad consisting of 2 real vectors $l^a$ and $n^a$, and two complex vectors $m^a$ and $\bar{m}^a$ (where the bar represents the complex conjugate). The null tetrad must satisfy the orthogonality conditions
\begin{equation}
l^an_a=-1;\qquad m^a\bar{m}_a=1,
\end{equation}
with all other inner products vanishing. One can show that the following null tetrad satisfies the above conditions
\begin{equation}
l^a = \frac{1}{\Delta(r)}\left(r^2+a^2, \Delta(r), 0, a\right)^a;
\end{equation}
\begin{equation}
n^a = \frac{1}{2(r^2+a^2\cos^2\theta)}\left(r^2+a^2, -\Delta(r), 0, a\right)^a;
\end{equation}
\begin{equation}
m^a = \frac{1}{\sqrt{2}(r+ia\cos\theta)}\left(ia\sin\theta, 0, 1, \frac{i}{\sin\theta}\right)^a;
\end{equation}
\begin{equation}
\bar{m}^a = \frac{1}{\sqrt{2}(r-ia\cos\theta)}\left(-ia\sin\theta, 0, 1, -\frac{i}{\sin\theta}\right)^a.
\end{equation}
Using this, the Teukolsky equation for the spacetime is given by
\begin{widetext}
\begin{equation} \label{E:Teukolsky_eqn}
\begin{gathered}
\mathcal{S}^P=\frac{1}{2}\left[\frac{(r^2+a^2)^2}{\Delta(r)}-a^2\sin^2\theta\right]\frac{\partial^2\Psi_0^P}{\partial t^2} + \left[-\frac{r^2+a^2}{2}\,x(r,\theta) + \frac{ia\sin\theta}{\sqrt{2}\,\bar{\rho}}\,y(r,\theta) - (r^2+a^2)\frac{\Delta'(r)}{\Delta(r)} + 4r + 2ia\cos\theta\right]\frac{\partial\Psi_0^P}{\partial t}\\
+ a\left[\frac{r^2+a^2}{\Delta(r)}-1\right]\frac{\partial^2\Psi_0^P}{\partial t\partial\phi} - \frac{\Delta(r)}{2}\,\frac{\partial^2\Psi_0^P}{\partial r^2} + \frac{\Delta(r)x(r,\theta)-3\Delta'(r)}{2}\,\frac{\partial\Psi_0^P}{\partial r} - \frac{1}{2}\,\frac{\partial^2\Psi_0^P}{\partial\theta^2} - \frac{1}{2}\left[\frac{\sqrt{2}\,y(r,\theta)}{\bar{\rho}}+\cot\theta\right]\frac{\partial\Psi_0^P}{\partial\theta}\\
+ \frac{1}{2}\left[\frac{a^2}{\Delta(r)}-\csc^2\theta\right]\frac{\partial^2\Psi_0^P}{\partial\phi^2} - \left[\frac{a}{2}\,x(r,\theta)-\frac{i\csc\theta}{\bar{\rho}}\,y(r,\theta) + a\frac{\Delta'(r)}{\Delta(r)} + 2i\csc\theta\cot\theta\right]\frac{\partial\Psi_0^P}{\partial\phi}\\
\left[\left(\Delta'(r)+\frac{3}{2}\,\rho\Delta(r)\right)x(r,\theta) + \frac{3\rho}{\sqrt{2}}\left[iar\sin\theta-\frac{\cot\theta}{3}(2(r^2+a^2)+a^2\sin^2\theta)\right]y(r,\theta)-\frac{5}{4}\,\Delta(r)+\frac{\cot^2\theta+3\csc^2\theta}{2}\right]\Psi_0^P,
\end{gathered}
\end{equation}
\end{widetext}
where 
\begin{equation}
x(r,\theta)=\frac{1}{\Psi_2}\left(2\rho\Phi_{11}^B+\frac{2r}{\Sigma}\,\Lambda^B - \frac{\Delta'''(r)}{24\Sigma}\right),
\end{equation}
\begin{equation}
y(r,\theta) = \frac{1}{\Psi_2}\left(-2\tau^B\Phi_{11}^B+\frac{\Delta(r)}{\Sigma^2}\left[\frac{\Delta'''(r)}{24}+2r\Lambda^B\right]\right),
\end{equation}
where $\Phi_{11}^B$ and $\Lambda^B$ are the two non-zero Newman--Penrose Ricci coefficients of the spacetime given by 
\begin{equation}
\Phi_{11}^B = \frac{4rf'(r)-\Sigma f''(r)-4f(r)}{8\Sigma^2},
\end{equation}
\begin{equation}
\Lambda^B = \frac{\Delta''(r)-2}{24\Sigma}\,,
\end{equation}
where
\begin{equation}
f(r)=\Delta(r)-r^2-a^2,
\end{equation}
and finally 
\begin{equation}
\rho = -\frac{1}{r-ia\cos\theta}\,.
\end{equation}
We now enforce separability of the RHS of the Teukolsky equation. That is, we specify 
\begin{equation}
\Psi^P_0 = e^{-i\omega t}e^{im\phi}R(r)S(\theta).
\end{equation}
If we insert this into equation \eqref{E:Teukolsky_eqn}, evaluate the derivatives acting on $\Psi_0$, divide by $\Psi_0$ and enforce separability, then we find that the following five conditions must hold
\begin{widetext}
\begin{equation} \label{E:T_eqn_constraints}
\begin{gathered}
\frac{\partial x(r,\theta)}{\partial\theta} = 0;\\
\frac{\partial y(r,\theta)}{dr}=0;\\
\frac{\partial^2}{\partial r\partial\theta}\left[-\frac{r^2+a^2}{2}\,x(r,\theta) + \frac{ia\sin\theta}{\sqrt{2}\,\bar{\rho}}\, y(r,\theta)\right] = 0;\\
\frac{\partial^2}{\partial r\partial\theta}\left[\frac{a}{2}\,x(r,\theta)+\frac{i\csc\theta}{\bar{\rho}}\,y(r,\theta)\right] = 0;\\
\frac{\partial^2}{\partial r\partial\theta}\left[\left(\Delta'(r)+\frac{3}{2}\,\rho\Delta(r)\right)x(r,\theta) + \frac{3\rho}{\sqrt{2}}\left[iar\sin\theta-\frac{\cot\theta}{3}(2(r^2+a^2)+a^2\sin^2\theta)\right]y(r,\theta)\right] = 0.
\end{gathered}
\end{equation}
\end{widetext}
For simplicity, let us define $u=r-ia\cos\theta=-1/\rho$, $W = f''u^2-6f'u+12f$ and finally $N=-u\Sigma f'''+2(ur+3\Sigma)f''-24tf'+24f$. Then we have 
\begin{equation}
\Psi_2 = \frac{W}{12\Sigma u^2}\,,
\end{equation}
and 
\begin{equation}
x(r,\theta) = \frac{N}{2\bar{u}W}\,.
\end{equation}
The first condition in equation \eqref{E:T_eqn_constraints} forces
\begin{equation} \label{E:N_condition}
N = \chi(r)\bar{u}W,
\end{equation}
where $\chi(r)$ is some function of $r$. Equation \eqref{E:N_condition} is a cubic polynomial in $\cos\theta$, if we expand both sides of the equation, we can then equate the terms at each order of $\cos\theta$. This gives us four equations
\begin{equation} \label{E:f_eqn_1}
f'''+\chi f'' =0;
\end{equation}
\begin{equation} \label{E:f_eqn_2}
6f''-rf'''+\chi(6f'-rf'')=0;
\end{equation}
\begin{equation} \label{E:f_eqn_3}
r^2f'''-2rf''+\chi(r^2f''-12f)=0;
\end{equation}
\begin{equation} \label{E:f_eqn_4}
24f-24rf'+8r^2f''-r^3f'''+\chi(6r^2f'-r^3f''-12rf)=0.
\end{equation}
If we substitute equation \eqref{E:f_eqn_1} into equation \eqref{E:f_eqn_2} we get
\begin{equation} \label{E:f_eqn_5}
f''=-\chi f'.
\end{equation}
If we substitute equation \eqref{E:f_eqn_1} into equation \eqref{E:f_eqn_3} we get
\begin{equation} \label{E:f_eqn_6}
rf''=-6\chi f.
\end{equation}
If we substitute equation \eqref{E:f_eqn_1} into equation \eqref{E:f_eqn_4} we get
\begin{equation} \label{E:f_eqn_7}
12(f-rf')=\chi r(rf'+6f).
\end{equation}
Now let us substitute equation \eqref{E:f_eqn_5} into equation \eqref{E:f_eqn_6}, then we find
\begin{equation} \label{E:f_eqn_8}
\chi(rf'-6f)=0.
\end{equation}
There are two possible cases, the first being $\chi\neq 0$. Then via equation \eqref{E:f_eqn_8}, the general solution is $f=c\, r^6$ where $c$ is some constant. Then via equation \eqref{E:f_eqn_5} we have that $\chi=-5/r$. But also via equation \eqref{E:f_eqn_1} we have $\chi=-4/r$. This is a contradiction unless $c=0$. But if this is the case then $f=0$ and hence $W=0$. This forces $\Psi_2$ to vanish, which was the only non-zero Weyl scalar. If all the Weyl scalars of the spacetime vanish, then the Weyl tensor is vanishing. Hence the spacetime is Petrov type O, that is, the spacetime is conformally flat. If we forbid this possibility, then we must enforce $\chi=0$. Then via equation \eqref{E:f_eqn_5} we find that the general solution of $f$ is $f=c\, r+b$ where $c$ and $b$ are constants. But via equation \eqref{E:f_eqn_7} we have $c\,r+b=c\,r$, hence $b=0$. So $f=c\,r$, that is, we have
\begin{equation}
\Delta(r)=r^2+a^2+c\,r.
\end{equation}
With this, one can check that the remaining four conditions in equation \eqref{E:T_eqn_constraints} are all satisfied. \\

Now in order to find the final unconstrained constant, $c$, we can find the effective Newtonian potential via
\begin{equation}
\Phi = \lim_{r\rightarrow\infty}-\frac{1}{2}(g_{tt}+1)=\frac{c}{2r}+\mathcal{O}\left(\frac{1}{r^3}\right).
\end{equation}
So in order to fit the Newtonian potential of a point mass at large $r$, we must choose $c=-2m$, where $m$ is the usual mass parameter. So after all of these constraints, we are left with the metric
\begin{widetext}
\begin{equation}
ds^2=-\left(1-\frac{2mr}{\rho^2}\right)dt^2-\frac{4mar\sin^2\theta}{\rho^2}\,dtd\phi + \frac{\rho^2}{\Delta}\,dr^2 + \rho^2d\theta^2 + \Sigma\sin^2\theta\, d\phi^2,
\end{equation}
\end{widetext}
where $\rho=\sqrt{r^2+a^2\cos^2\theta}$, $\Delta=r^2+a^2-2mr$ and $\Sigma=r^2+a^2+2mra^2\sin^2\theta/\rho^2$. This is just the Kerr solution as written in Boyer--Lindquist coordinates, and hence concludes the proof. \hfill $\Box$\\

Note that some of the conditions of the theorem are degenerate. The existence of a Killing--Yano tensor automatically ensures that the Hamilton--Jacobi equation and the Klein--Gordon equation are separable. So while the original form of the theorem may be better suited for providing a list of conditions which can potentially be linked to physical observations that can be checked off to determine to applicability of the theorem to real astrophysical black holes, mathematically speaking, we can write the theorem in a shorter form.\\

\noindent\emph{Theorem:} \\
\enlargethispage{20pt}
\noindent Suppose a 4-dimensional spacetime satisfies the following conditions:
\begin{enumerate}
    \item The spacetime is stationary and axisymmetric
    \item The spacetime admits a Killing--Yano tensor
    \item The spacetime admits a separable Teukolsky equation
    \item The spacetime is asymptotically flat
    \item The spacetime is not conformally flat
    \item The spacetime reproduces the Newtonian gravitational potential of a point mass at sufficiently large $r$
\end{enumerate}
Then the spacetime is uniquely given by the Kerr solution. \\

\noindent\emph{Conclusions:}

We have shown that, under a certain set of conditions as stated above, the metric of a spacetime is uniquely given by the Kerr metric. Crucially, none of these conditions assumes the Einstein equations. This distinguishes our result from the classical uniqueness theorems of Israel, Carter, Robinson, and Hawking, all of which a priori assume that the vacuum Einstein field equations hold. Although the spacetime which results from the theorem presented in this paper is Ricci flat, this is not explicitly (nor implicitly) an assumed condition of the theorem, rather it emerges as a natural consequence of the stated conditions.\\

This has direct implications for quantum gravity and modified theories of gravity. Any such theory which admits regular black holes must violate at least one condition of our theorem. On the other hand, should observational evidence accumulate in favour of each condition of our theorem, one would be compelled to conclude that the spacetime of real astrophysical black holes in our universe is described by the Kerr solution, and moreover the vacuum Einstein equations will be satisfied.\\

Our result carries a further consequence for the existence of singularities. Because the theorem forces the spacetime to be exactly Kerr, the presence of a singularity follows immediately, without any appeal to the field equations. This establishes the presence of a singularity by a route complementary to the Penrose singularity theorem. Where Penrose's result invokes the Einstein equations within a far more general dynamical setting, ours shows that, under the specific conditions considered here, a singularity is unavoidable even when the validity of the Einstein equations is not presupposed.\\

\emph{Acknowledgements:}
The author would like to thank Prof. Andrew Melatos for proof reading the manuscript. This research was conducted by the Australian Research Council Centre of Excellence for Gravitational Wave Discovery (OzGrav), through project number CE230100016.



\end{document}